\pdfoutput=1
\documentclass[12]{article}

\usepackage{mathptmx}

\usepackage[latin9]{inputenc}
\usepackage{textcomp}
\usepackage{amsmath}
\usepackage{amssymb}
\usepackage{graphicx}
\usepackage[authoryear]{natbib}
\usepackage{wasysym}
\usepackage{url}
\usepackage{lscape,float}
\usepackage{xcolor}
\usepackage{caption}

\usepackage{amsfonts}
\begin{document}
\baselineskip=12pt 

\begin{center}
\vspace{1.5cm}
 {\huge Efficient Bayesian species tree inference under \vspace{.4cm}\\
 the multi-species coalescent}\\
{\huge{} \vspace{1.2cm}
 }{\Large Bruce Rannala$^{1}$ and Ziheng Yang$^{2}$}\\
 \vspace{1.2cm}
\par\end{center}

\begin{center}
$^{1}$Department of Evolution \& Ecology, University of California, Davis, CA 95616, USA \\
 \vspace{0.2cm}
$^{2}$Department of Genetics, Evolution and Environment, University College London, London WC1E 6BT, UK \\
 \vspace{0.2cm}
 \par\end{center}

\vspace{1cm}
  \vspace{1.2cm}

\noindent \textbf{Correspondence to:} \\
Ziheng Yang \\
Department of Genetics, Evolution and Environment, University College London, London WC1E 6BT, UK  \\
\textbf{Phone:} 020 7679 4379 \\
\textbf{Email:} z.yang@ucl.ac.uk \vspace{1cm}

\noindent \textbf{Running Head:} Bayesian species tree inference \\
\vspace{1cm}

\newpage
\noindent
\section*{Abstract}
A method was developed for Bayesian inference of species
phylogeny using the multi-species coalescent model.
To improve the mixing properties of the Markov chain Monte Carlo (MCMC) algorithm that traverses the space of species trees, we implement two efficient
MCMC proposals: the first is based on the Subtree Pruning and Regrafting (SPR) algorithm and the second
is based on a novel node-slider algorithm. Like the Nearest-Neighbor Interchange (NNI) algorithm we implemented
previously, both algorithms propose changes to the species tree, while simultaneously
altering the gene trees at multiple genetic loci to automatically avoid conflicts with the newly-proposed species tree.
The method integrates over gene trees, naturally taking account of the uncertainty of gene
tree topology and branch lengths given the sequence data.
A simulation study was performed to examine the statistical properties of the new method.
We found that it has excellent statistical performance, inferring the correct species tree
with near certainty when analyzing 10 loci.  The prior on species trees has some impact, particularly
for small numbers of loci.  An empirical dataset (for rattlesnakes) was reanalyzed. While the
18 nuclear loci and one mitochondrial locus support largely consistent species trees under
the multi-species coalescent model estimates of parameters suggest drastically different evolutionary
dynamics between the nuclear and mitochondrial loci.

\section*{Introduction}
Multilocus genetic sequence data have gained importance in inferring
species trees in recent years and several inference methods have been
proposed for this purpose. As noted by \cite{Maddison:1997} several processes
can cause the species tree to differ from gene
trees underlying particular loci.  Some of these processes, such as
introgression between species and horizontal gene transfer, involve
reticulations in the species tree, whereas others, such as incomplete lineage
sorting and gene duplications, occur within the context of a
non-reticulate (and typically binary) species tree. An important
potential source of gene-tree versus species-tree conflicts among
genetically isolated species is incomplete lineage sorting; this
phenomenon is typically modeled using a coalescence process.  A simple
widely-used method for multilocus species tree inference concatenates
sequences from different loci, assuming that a single tree (treated as
the species tree) underlies all the loci \cite[reviewed
in][]{Rannala:2008, Edwards:2009}. This approach can lead to strongly supported
incorrect phylogenetic trees when incomplete lineage sorting occurs
\cite[see e.g.,][]{Leache:2011}, and has been shown to be inconsistent
\cite[]{Kubatko:2007}.  Another heuristic
approach is to infer separate gene trees and then attempt to reconcile
the differences among gene trees to obtain an estimate of the species
tree \cite[][]{Page:1997}.  Use of the most frequent gene tree among loci as the
species tree estimate can be inconsistent  in the so-called 'anomaly zone' \cite[]{Degnan:2005, Degnan:2006}.

\cite{Maddison:1997} and \cite{Maddison:2006} proposed a parsimony-inspired
method for inferring the species tree called minimizing deep
coalescence (MDC) events for gene trees.  Other examples include
species tree estimation by
minimizing coalescence times across genes \cite[the Global LAteSt
Split, GLASS;][]{Mossel:2010}, by using the
average ranks of coalescences \cite[STAR;][]{Liu:2009}, by
using average coalescence times \cite[STEAC; ][]{Liu:2009},
by using maximum likelihood for gene trees under
coalescence \cite[STEM;][]{Kubatko:2009} and by maximum pseudo-likelihood
\cite[M-PEST;][]{Liu:2010}.  These approximations treat the estimated
gene trees (including either the gene tree topology alone or both the
gene tree topology and branch lengths) as observations (data),
ignoring phylogenetic uncertainties.  Such approximations can lead to
systematic biases as well as underestimation of the uncertainty of
inferred species trees \cite[][]{Leache:2011}.

A parametric statistical method for inferring the species tree using multi-locus
sequence data should integrate over the (unobserved) gene trees (both
the tree topology and branch lengths).  For the case of three species,
with one sequence from each species at each locus, a maximum
likelihood method was implemented by \cite{Yang:2002}, using numerical
integration to integrate out the two coalescent times in each gene
tree.  For larger problems with more species or more sequences,
maximum likelihood is not computationally feasible.  Instead the
Bayesian method is used, with Markov chain Monte Carlo (MCMC) used for
the computation.  A few MCMC implementations now exist to estimate species
trees \cite[][]{Edwards:2007, Heled:2010, Ronquist:2012} although they
are limited to a small number of species and loci.

The mutual constraints imposed by the gene trees and the species tree are a challenge for developing
an efficient MCMC inference program under the multi-species coalescent model.
For example, if the divergence time between species $A$ and $B$  is
$\tau_{AB}$, the divergence time between two sequences from $A$ and $B$
must be older with $t_{AB} > \tau_{AB}$ at every locus (sequences must
diverge before the species diverge). Such constraints can cause serious difficulties
leading to poor MCMC mixing when one attempts to change species
divergence times and/or the species tree topology if the gene trees are fixed.
There are two possible solutions to this difficulty: either analytically integrate
over the gene trees (rather than performing MCMC) preserving the constraints; or develop efficient MCMC
proposals for jointly modifying the species tree and the gene trees that
obey the constraints.
Recent methods that have been
developed for inferring species trees from single nucleotide
polymorphism (SNP) data follow the first strategy \cite[][]{Bryant:2012}.  The simplicity of
these data allow the gene trees to be analytically integrated out of
the model. However, a drawback of such methods is that SNPs provide
very little information about branch lengths in the gene trees and the
power is therefore expected to be reduced in comparison with
sequence-based methods.

Here, we follow the second approach in developing a Bayesian inference procedure for
the analysis of multi-locus sequence data that jointly infers the
species phylogeny and gene trees as well as other
relevant parameters such as ancestral population sizes and species
divergence times.  We extend our program BPP (for Bayesian
Phylogenetics and Phylogeography)
\cite[][]{Yang:2010hm,Rannala:2013gk,Yang:2014b} to allow this joint
inference. We develop two novel MCMC proposals to change the species
tree, at the same time modifying the gene trees to avoid conficts
between the gene trees and newly proposed species tree.  The first
move is based on the Subtree Pruning and Regrafting (SPR) algorithm
for rooted trees.  This changes the species tree topology while
preserving the node ages in the species tree as well as in the gene
trees.  The second move is based on a node-slider algorithm, which
changes the topology as well as the node ages in the species tree and
gene trees. The two new proposal algorithms lead to greatly improved
mixing behavior of the MCMC by comparison with the simple NNI
algorithm implemented in our previous work \cite[]{Yang:2014b}.

\section*{Theory}
Here we review the formulation of the species tree inference problem
in a Bayesian framework and then describe our new MCMC
algorithms.  Let $X_i$ be the sequence alignment for locus $i$. The
number of sequences per species may vary for each locus and some
species may not be sampled for a particular locus.  Our
requirement is that every locus should have at least two sequences.
Let there be $L$ loci and define $\mathbf{X} = \{ X_i \}$ to be the
full dataset. Let $G_i$ be the gene tree for the sequences sampled at
locus $i$ (including both the gene tree topology and branch lengths or
coalescent times).  Let $\mathbf{G} = \{ G_i \}$.  We assume the loci
are unlinked so that
\begin{equation}
f(\mathbf{X}|\mathbf{G},\omega) = \prod_{i=1}^L f(X_i | G_i,\omega).
\end{equation}
where $f(X_i|G_i,\omega)$ is the {\em phylogenetic likelihood} calculated according to the usual Felsenstein algorithm \cite[]{Felsenstein:1981}
and $\omega$ is a vector of parameters in the nucleotide substitution model.
The posterior probability of the species tree ($S$) and the parameters is given by
\begin{equation}
f(S,\Theta|\mathbf{X}) = \frac{1}{f(\mathbf{X})}\int_{\omega}\int_{G} f(\mathbf{X}|\mathbf{G},\omega)f(\mathbf{G} | S, \Theta) f(\omega) f(\Theta) f(S) \ \textrm{d}\mathbf{G} \textrm{d}\omega,
\end{equation}
where $\Theta = \{ \{\tau_j\}, \{ \theta_j \} \}$ is the set of
$\theta$s and $\tau$s associated with the species tree(s).  Note that
$\theta_j = 4N_j \mu$ where $N_j$ is the effective population size of
(ancestral or contemporary) species $j$. The branch lengths in the
gene trees and the divergence times in the species trees are
represented in units of expected number of substitutions per site \cite[as in
][]{Rannala:2003}. The term $f(\mathbf{G} | S, \Theta)$ is the
multispecies coalescent density of gene trees (topology and coalescent times)
given the species tree \cite[][]{Rannala:2003}.  We use MCMC
to generate a sample from the posterior density
of $S$ and $\Theta$.  Here we focus on two new MCMC proposals that
efficiently propose changes to the species tree topology ($S$).  The moves
that do not alter the species tree topology are identical to those
described in \cite{Rannala:2003, Rannala:2013gk}.  The first move, based on the SPR
algorithm, is a direct extension of the Nearest-Neighbor Interchange (NNI)
algorithm implemented in \cite{Yang:2014b}.  The second move, based
on a node-slider algorithm, changes the topology as well as a node
age in the species tree.

\subsection*{SPR algorithm to modify the species tree topology}
Let anc($a$) be the mother node of node $a$.  We refer to the branch
anc($a$)-$a$ as branch $a$.  We define clade
or subtree $a$ to include $a$, all its descendents, and branch $a$.
Nodes on the species tree are represented by capital letters, such as
$A$, and their ages are denoted by $\tau$s (such as $\tau_A$).  Nodes on
gene trees are labeled using small-case letters, and their ages are
denoted by $t$s.

Our SPR move prunes off the branch $Y$-$A$ (including clade $A$) and
reattaches it to a target branch $C$, retaining the same age $\tau_Y$ at reattachment (fig.~1).  Our
algorithm does not change divergence times ($\tau$s) in the species
tree or node ages ($t$s) in the gene trees.  We preferentially propose
changes to the species tree topology around short (rather than long) internal branches.
We sample an internal branch $i$ (out of $s-2$ internal branches for a species tree of
$s$ species) according to the following probabilities
\begin{equation}\label{BranchWeight}
   w_i \propto b_{i}^{-\frac{1}{2}},
\end{equation}
where $b_i$ is the length of the internal branch.  The sampled branch
is branch $X$-$Y$.  Node $Y$ has two daughter branches.  We sample one
at random and let it be $A$; the other will be $B$.  We then prune off
branch $Y$-$A$ (including clade $A$) and reattach it to branch $C$
in the species tree.  Let $Z$ be the most recent common ancestor of
$A$ and $C$, with age $\tau_Z$.  The move affects species on the path
$A$-$Z$-$C$. For the SPR move illustrated in fig.~1, $Y$ is species $AB$,
$X$ is $ABD$, and $Z$ is $ABCD$.

Among the feasible target branches for reattachment, we sample one using a
probability distribution that favors small changes to the species tree topology.  A feasible target
branch is a branch of the species tree that remains after branch $Y$-$A$ is
pruned off (exclusive of branch $B$) and that covers the age $\tau_Y$ (see fig.~1).  In
choosing a target branch, we use probabilities
\begin{equation} \label{TargetWeight}
    v_i \propto 1/c_i,
\end{equation}
where $c_i$ is the number of nodes on the path $A$-$Z$-$C_i$ for
potential target branch $C_i$.  The minimum for $c_i$ is 3, in which case $Z
= X$ and the SPR move reduces to the NNI move of \cite{Yang:2014b}.  Our proposal using
equation \ref{TargetWeight} thus favours small changes to the species
tree topology.

The move affects nodes on the gene trees that have age $\tau_Y < t <
\tau_Z$.  A \emph{moved} node (marked with $\bullet$ in fig.~1) lies in species
$AB$ ($Y$) or another ancestral species on the path from $Y$ to $Z$
(excluding $Z$ itself) and has exactly one daughter node with
descendants in $A$ only. The other daughter node has descendants in
one or more non-$A$ descendent populations as well.  The moved node
(and the descendant clade) is pruned and re-grafted to a randomly
chosen contemporary branch of the gene tree residing in a species on
the path from $C$ to $Z$.  In addition, four other kinds of
\emph{affected} nodes have their population IDs changed. Any node marked
with $\bigcirc$ or $\bigtriangleup$ has descendents in species $A$
only and changes its population ID from $AB$ ($Y$) to $AC$.  Any node
marked with $\Diamond$ is in species $C$ with age between
$\tau_Y$ and $\tau_{anc(C)}$ and changes its population ID from $C$ to
$AC$.  Any node marked with $\square$ is in species $AB$ with both
 daughter nodes having descendants in species $B$, and changes
its population ID from $AB$ to $B$. The proposal ratio incurred by the
move can be easily derived using a procedure similar to that used for the
NNI move \cite[]{Yang:2014b}

\subsection*{Nodeslider algorithm to modify the species tree}
\subsubsection*{Overview of the algorithm}
The nodeslider move prunes
off branch $Y$-$A$ (including clade $A$) in the species tree,
changes $\tau_Y$ and rescales the ages inside the $A$ clade
proportionally, and then reattaches the branch (and the $A$ clade) to a
new branch $C$ in the remaining species tree (fig.~2).  This proposal consists
of a pair of opposing moves, referred to as the `Expand' and
`Shrink' moves.  In the Expand move (towards the root), $\tau_Y$
increases, and the target branch $C$ is ancestral to node $Y$.  In the
Shrink move (towards the tips), $\tau_Y$ decreases, and the
target branch $C$ is one of the descendents of $B$, the sibling of $A$.
Thus the move slides node $Y$ and the attached clade $A$ either
towards the root, with the node ages in subtree $A$ expanded, or to
a descendent branch of the sibling species $B$, with the node ages in
subtree $A$ shrunk.

\subsubsection*{Details of the algorithm}
\noindent
{\bf 1.~Changes to the species tree: }
We describe the changes to the species tree
first.  A uniform random variable $U$ on $(0,1)$ is generated to decide whether to expand ($U \geq 0.5$) or shrink ($U < 0.5$).
In the Expand move (change from $S$ to $S^*$ in fig.~2), we use equation \ref{BranchWeight} to sample an
internal branch (out of $s - 2$) on the species tree and let it be
$X$-$Y$.  Node $Y$ has two daughter nodes.  We sample one at random
and let it be $A$; the other will be $B$.
In the example of fig.~2, $X$ is node 6, and $Y$ is node 7.
We then propose a new age
$\tau_Y^*$ for node $Y$ using an exponential density,
\begin{equation}
f(\tau_Y^*) = \left(\frac{1}{0.1\tau_X}\right)
\mathrm{e}^{-\frac{1}{0.1\tau_X} (\tau_Y^* - \tau_X)}, \ \ \ \tau_X < \tau_Y^* < \infty.
\end{equation}
In other words, the excess $\tau_Y^* - \tau_X$ has mean
$0.1\tau_X$.  The value $0.1$ is the ``Expand ratio'' and is
adjustable; we suspect small values close to zero are preferable.
Branch $Y$-$A$ (including clade $A$) is pruned off and reattached to the
remaining species tree at age $\tau_Y^*$.  There will be only one
ancetral branch which covers the new age $\tau_Y^*$.  Note that this
can be the root, in which case node $Y$ will become the new root.

In the Shrink move, we use equation \ref{BranchWeight} to sample an
internal branch on the species tree and let it be $Y$-$B$.  The other
daughter of node $Y$ (that is, the sibling of $B$) will be $A$.
Branch $Y$-$A$ (including clade $A$) will be pruned off and reattached
to a target branch $C$ that is a descendent of species $B$ at time $\tau_Y^*$.  The
new age $\tau_Y^*$ is proposed using a power density,
\begin{equation}
f(\tau_Y^*) = \frac{\lambda}{\tau_{B}} \left(\frac{\tau_Y^*}{\tau_B}\right)^{\lambda-1}, \ \ \ 0 < \tau_Y^* < \tau_B.
\end{equation}
This is a uniform density if $\lambda = 1$.  To simulate from the
power density we use the inverse transformation method. Generate a
uniform random variable $u \sim U(0, 1)$ and set
\begin{equation}
\tau_Y^* = \tau_B \times u^{1/\lambda}.
\end{equation}
We choose $\lambda=\log(0.1)/\log(0.9) = 21.85$ so that 90\% of the
density is within 10\% of $\tau_B$, with $\tau_Y^* \geq 0.9 \times
\tau_B$.  Here the value 10\% is called the ``Shrink ratio.'' We favor small values
like 10\% so that the new $\tau_Y^*$, smaller than $\tau_B$,
tends to be close to it.

When branch $Y$-$A$ is pruned off, the ages of all daughter nodes of
$A$ are rescaled by the factor $\tau_Y^* / \tau_Y$.  In the Expand
move, we slide branch $Y$-$A$ (including clade $A$) towards the root.
We find the branch ancestral to $X$ at age $\tau_Y^*$ for
reattachment.  In the Shrink move, we select a branch at random as
target for reattachment, among all branches that are descendents of
$B$ and exist at time point $\tau_Y^*$.  The number of branches for
reattachment is 1 in the Expand move, and is greater than 1 in the
Shrink move.

We now consider the factor in the proposal ratio incurred by changes
to the species tree.  First, consider the Shrink move (e.g., $S^* \rightarrow S$ in fig.~2).
Let $G(\tau_{Y}^*)$ be the number of
branches existing at time $\tau_Y^*$ that are descendents of node
$B$. Each of the branches is chosen as the grafting target of the
moved branch ($Y$-$A$) with probability $1/G(\tau_Y^*)$.  Furthermore,
if the move changes the age of the root ($\tau_0$) on the species
tree, the prior on the node ages in the species tree (the $\tau$s)
will influence the proposal ratio \cite[see equation 2 in][]{Yang:2010hm}.  The
factor in the proposal ratio due to the changes to the species tree is
thus
\begin{equation}
R_{\textrm{Shrink}} = \frac{w_{Y^*}}{w_Y} \times \frac{0.5}{1} \times G(\tau_Y^*)
\times \frac{\left(\frac{1}{0.1\tau_X^*}\right) \mathrm{e}^{-\frac{1}{0.1\tau_X^*} (\tau_Y - \tau_X^*)}}{\frac{\lambda}{\tau_B}\left(\frac{\tau_Y^*}{\tau_B}\right)^{\lambda-1}}
\times (\frac{\tau_Y^*}{\tau_Y})^m
\times \frac{g(\tau_0^*)}{g(\tau_0)} \times \left(\frac{\tau_0^*}{\tau_0}\right)^{-(s - 2)},
\end{equation}
where $m$ is the number of daughter nodes of $A$ whose ages are
rescaled, $\tau_X^*$ is the age of the mother node of $Y^*$, $g$ is
the gamma prior density for the root age $\tau_0$, and $s-2$ is the
number of non-root interior nodes.  The probability 0.5 is due to the
sampling to choose $A$ in the reverse Expand move.  For the Expand
move, the factor in the proposal ratio is $R_{\textrm{Expand}} =
1/R_{\textrm{Shrink}}$.

\noindent
{\bf 2. Changes to the gene trees:}  The gene trees are modified to avoid conflicts with the newly proposed species tree, similarly to the SPR algorithm.  Some nodes are pruned off the gene tree and regrafted back and some nodes have their population IDs changed due to the disappearance and appearance of populations.
In the Expand move, the target population $C$ is ancestral to $A$,
$B$, and $AB$ ($Y$).  In the Shrink move, the target population $C$ is
a descendent of $B$ (see fig.~2). We scan the gene tree at each
locus to identify the moved nodes.  A moved node (marked with
$\bullet$ in fig.~2) has exactly one daughter node with descendents in $A$
only.  We prune off the moved node, rescale the ages inside the clade
by $\tau_Y^* / \tau_Y$ and re-graft it to a randomly-chosen branch
that exists at the new time $t^* = t \times \frac{\tau_Y^*}{\tau_Y}$.
Note that there may be multiple target gene-tree branches.

At every locus, there may be multiple moved nodes and thus multiple
subtree pruning and regrafting operations on the gene trees.  These are thus conducted
in a disciplined manner, as follows.  We prune off all moved nodes
(and their attached subtrees), and ``lay them on the ground''.  We then
determine the new ages (rescaled by $\tau_Y^* / \tau_Y$) at which the
pruned subtrees are reattached, and identify and mark the reattachment
points.  The remaining part of the gene tree, called the skeleton, is
not changed (black branches in the gene trees of fig.~2) except that gene tree nodes
in population $AB$ become nodes in population $B$ and nodes in
population $C$ with age greater than $\tau_Y^*$ become nodes in
population $AC$ (fig.~2).   We then reattach the
pruned branches to the gene tree.  The order of pruning and
reattachment of the affected nodes is thus inconsequential.  In this
way we do not allow regrafting of one pruned branch onto another
pruned branch, but it may be possible for a pruned branch to be
regrafted to the same branch on the skeleton, so that the operation
may change the node ages without changing the gene tree topology. It
is also possible for multiple pruned subtrees to be reattached to the
same branch of the skeleton (at different time points).

If all sequences at a locus are from populations in clade $A$ on the species tree, all node ages on
the gene tree are rescaled by $\tau_Y^* / \tau_Y$, while their
population IDs remain unchanged.  This rescaling is necessary as
otherwise the gene tree may be in conflict with the proposed new
species tree.

The changes to the gene trees will incur a factor in the proposal
ratio, because the following components may not be the same in the
forward and reverse moves: the number of target branches for
reattaching the moved node, the probability density of the gene tree
given the species tree topology and parameters $\tau$s and $\theta$s
(i.e., the multispecies coalescent density), the rescaling of all the
gene-tree node ages (by the factor $\tau_Y^* / \tau_Y$), as well as
the probability of the sequence alignment given the gene tree at each
locus (the phylogenetic likelihood).

In the case of only three species, the nodeslider move reduces to a variant of the general NNI algorithm for rooted trees \cite[][p.293]{Yang:2014}, although it differs from the NNI algorithm implemented by \cite{Yang:2014b} or the SPR move described above.  Here the nodeslider move changes both the species tree topology and a species divergence time ($\tau$), and always changes the root of the species tree (fig.\ 3).  In the Expand move (from $S \rightarrow S^*$ in fig.\ 3), branch $Y$-$A$ is pruned off, the age $\tau_Y$ is increased to $\tau^*_Y > \tau_X$, and the branch is reattached to the species tree, with node $Y$ becoming the new root.  The reverse Shrink move slides the root of the species tree towards the tips so that the younger node becomes the new root.

We note that each of the three moves that we have implemented to change the species tree topology, including the NNI of  \cite{Yang:2014b}  and the SPR and nodeslider moves of this paper, is sufficient to allow the MCMC to traverse the whole space of the species trees.  In BPP, we use SPR (which includes NNI as a special case) and nodeslider moves with pre-assigned probabilities (such as 0.6 for SPR and 0.4 for nodeslider).

\subsection*{Validation of the theory and implementation}
The new SPR and nodeslider moves are implemented in BPP version~3.2.
Our algorithms are complex and extensive testing has been conducted to
confirm the correctness of the theory and the implementation.  Because our new
moves do not affect the calculation of the phylogenetic likelihood our tests have focused mainly on
generating the prior for the species trees and parameters of the multi-species coalescent
model ($\theta$s and $\tau$s) via MCMC when the sequence likelihood is fixed at 1.
We confirmed that the SPR and nodeslider algorithms, used either alone or in
combination, sampled the species trees correctly according to the prior, which is
analytically available for four different priors described by \cite{Yang:2014b} and \cite{Yang:2015}
for the cases of 3, 4 and 5 species.

\subsection*{Summary of the posterior}
The BPP program generates an MCMC sample from the posterior probability distribution of species phylogenies
and the posterior distribution of parameters ($\tau$s and $\theta$s) given each phylogeny.  Here we
focus on summaries of the species trees.  For species
tree inference the model with the highest posterior probability is presented as the
maximum \textit{a posteriori} (MAP) model.  The program calculates support values for clades on the MAP tree.
The program also generates posterior probabilities for
individual clades as well as the majority-rule consensus tree, with support values.
If the model
parameters  ($\theta$s and $\tau$s) are of primary interest we suggest that one should run the program
a second time with the species tree fixed to the MAP tree or concensus tree \cite[analysis A00,][]{Yang:2015}.
For an example see fig.~6 (below) generated by our empirical analysis of the rattlesnake dataset
of  \cite{Kubatko:2011}.

\section*{Results}

\subsection*{Simulation analysis of statistical performance}
Simulations were used to examine the influence
on the posterior probabilities of species trees of
the number of loci, the mutation rate (sequence divergence level), and the
prior on topology.  We simulated data under the
multispecies coalescent model using either a completely symmetrical
or asymmetrical tree of 16 species with two sequences
sampled per species (32 sequences in total) (see~fig.~4). For
simplicity, we assumed equal $\theta$s among ancestral and
contemporary species with either $\theta=0.001$ (low mutation rate) or $\theta=0.01$ (high mutation rate). We
set all internal branch lengths equal to $\theta$, so that $\tau_i -
\tau_j = \theta$ where node $i$ is the mother of node $j$. Thus, the
height of the root was $\tau_0 = 15 \times \theta$ for the asymmetrical tree
and $\tau_0 = 4 \times \theta$ for the symmetrical tree.
For each of the $2 \times 2 = 4$ parameter/topology combinations 50
datasets were simulated of either $L=2$ or $L=10$ unlinked
loci, each with $n=1000$ sites. Thus, $2 \times 2 \times 2 \times 50 =
400$ datasets were simulated in total.
The MCcoal program which is part of the BPP package was used to
generate gene trees under the multi-species coalescent model
and to simulate sequence alignments on the trees  under the JC69 model.

The simulated
datasets were analyzed using the BPP 3.2 program with a $G(2, 200)$
prior for $\theta$ when the true $\theta=0.01$ and a $G(2, 2000)$
prior for $\theta$ when the true $\theta=0.001$.  While the prior
means match the true values, the gamma distribution with shape
parameter 2 is diffuse (uninformative).
Similarly, a gamma prior with shape parameter 2 and with mean equal
to the true value was assigned to $\tau_0$, the age of the root of the species tree. In other words,
the prior on $\tau_0$ was $G(2, 50)$ for datasets simulated under
a symmetrical tree with $\theta=0.01$, $G(2, 13.3)$ for the asymmetrical tree with
$\theta=0.01$, $G(2, 500)$ for the symmetrical tree with $\theta=0.001$,
and $G(2, 133)$ for datasets simulated under an asymmetrical tree with $\theta=0.001$.
Two analyses were carried out for each dataset using
different priors on the tree topology: a uniform prior on labeled histories
(with rank-ordered nodes) and a uniform prior on rooted trees (ignoring node ordering). Each of the simulated datasets
(and prior combinations) were analyzed using two independent MCMC runs
with different starting seeds to check convergence.  Thus $400 \times
2 \times 2 = 1600$ MCMC runs were carried out in total. Each MCMC
analysis was run for 200,000 iterations, sampling every second
iteration and discarding the first 50,000 iterations as burn-in.

To examine the statistical performance of the method we calculated the
proportion of datasets (among 50 replicate simulations) in which each of the 14 nodes in the true species
tree is found in the consensus tree; note that a node of the true tree is in the
consensus tree if its posterior probability is $>0.5$. This is a measure of power.
We also examined the empirical coverage
of the $95\%$ and $99\%$ Credible Set of Trees (CST).  The empirical
coverage is defined as the proportion of credible sets that contain the true tree.
The results are summarized in Table~1.
The method performs very well in identifying the true clades, even with only 2 loci. With the exception of nodes 12 to 14
at the base of the tree (see fig.~4) all nodes of the true tree are present in the
concensus tree with frequencies of 0.76 or greater. The empirical coverage of the credible set of trees
provides a measure of the accuracy of the method. The accuracy is very high, with the true tree contained in both the 95\%
and 99 \% credible sets of trees for $100\%$ of simulated data analyses in all but two cases:
(1) trees inferred using the uniform labeled history prior
from the data simulated on an asymmetrical tree with 2 loci and with $\theta=0.001$ -- the coverage is 0.92 for the
$95\%$ and  $99\%$ CSTs; and
(2) trees inferred using the uniform topology prior from the data simulated on a symmetrical tree with 2 loci and with $\theta=0.001$
-- the coverage is 0.98 for the $95\%$ CST and 1.0 for the $99\%$ CST. In other words,
in all but one case the coverage is greater than the nominal value of either $95\%$ or $99\%$.

The mean number of trees contained in the $99\%$ CST provides a measure of the precision of the estimator
of species tree topology (fig.~5 and Table~1). The mean number of trees ranged from a minimum of $1.1$ (for 10 loci, $\theta=0.01$ and a symmetrical
true tree and a uniform prior on labeled histories) to a maximum of $6297.3$  (for 2 loci, $\theta=0.001$ and a symmetrical true tree with a uniform topology
prior). It is evident from the mean number of trees in the CST that the prior on topology can have a large effect on
the precision of the method (fig.~5 and Table~1). The uniform prior on labeled histories tends to favor symmetrical over asymmetrical trees while the
uniform prior on topologies increases the probability of asymmetrical trees.  When the prior favors the shape of the true tree,
the estimates are more precise with a smaller CST. For example, the histograms in rows 1 and 3 of fig.~5 which represent analyses
of symmetrical (Sym) trees, show an increase in the variance and mean of the number of trees in the 99\% CST when using
a uniform prior on rooted trees (tree) versus a uniform prior on labeled histories (LH). Conversely, the variance and mean decrease
when analyzing data simulated on asymmetrical (Asym) trees as seen in rows 2 and 4 of fig.~5.
In other words,
the number of trees in the credible set is increased when the uniform prior on rooted trees is used to analyze data generated using a symmetrical tree.
Conversely, the number of trees in the credible set is increased when a  uniform labeled history prior is used to analyze data generated using
an asymmetrical tree. The influence of the prior decreases when the number of loci increases from 2 to 10 and is negligible for
the informative data simulated using $\theta=0.01$ (see Table~1 and fig.~5).

\subsection*{Analysis of empirical datasets}
The empirical dataset we analyze here includes 18 nuclear loci from
six subspecies of \emph{Sistrurus} rattlesnakes, generated and
analyzed by \cite{Kubatko:2011}.  Rattlesnakes are venomous snakes of
the New World, with species falling into two genera: \emph{Crotalus}
which contains more than 20 species and \emph{Sistrurus} which contains
three named species: \emph{catenatus}, \emph{miliarius}, and
\emph{ravus}.  However, mtDNA suggests that \emph{ravus} in fact
belongs to the genus \emph{Crotalus} \cite[]{Murphy:2002,Parkinson:2002}.
The data analyzed here are from \emph{S. catenatus} and
\emph{S. miliarius} only.  Within each of these two species, three subspecies
are formally described on the basis of morphological variation in
scale characters, body size and coloration, and geographic
distribution.
The three \emph{S. catenatus} subspecies are
\emph{S.~c.~catenatus} (C), \emph{S.~c.~tergeminus} (T), and
\emph{S.~c.~edwardsii} (E), while the three \emph{S.~miliarius}
subspecies are \emph{S.~m.~miliarius} (M), \emph{S.~m.~barbouri} (B),
and \emph{S.~m.~streckeri} (S).
The data also include sequences from
two outgroup species: \emph{Agkistrodon contortrix} (Ac) and
\emph{A. piscivorus} (Ap).  Although the BPP analysis does not require
outgroups and those two outgroup species appear quite distant from the
ingroup species, we use them as well for easy comparison with the
results of \cite{Kubatko:2011}.  We analyze the 18 nuclear loci and the
single mitochondrial locus separately, since they have very different
characterizations, including different
mutation rates and effective population sizes.

\noindent
{\bf Analysis of the nuclear loci:}  Among the 18 loci, the number of
sequences per locus ranges from 48 to 52, and the sequence length
ranges from 194 to 849 \cite[][table 2]{Kubatko:2011}.  We use the uniform prior for rooted trees
\cite[]{Yang:2014b}.  For the parameters on the species tree, we use the
prior $\theta \sim G(2, 1000)$ with the prior mean 0.002 (2
differences per kb), and $\tau_0 \sim G(1.2, 100)$ with the prior mean
for the age of the root
to be 1.2\%.  The shape parameters (2 and 1.2) mean that
the gamma priors are diffuse, while the means are chosen to be
plausible for the data, based on preliminary runs of the A00 analysis
({\tt speciesdelimitation=0, speciestree=0})
under a reasonable tree \cite[]{Yang:2015}.

We use 8000 iterations for the burnin, after which we take $2 \times 10^5$
samples, sampling every 4 iterations.  We run each analysis twice to
check for consistency between runs and then merge the samples to
produce summary results.  Each run took about 10 hours on one CPU core.
\cite{Kubatko:2011} reported running times of  $\sim$10 days using StarBEAST,
suggesting that the algorithms in BPP may produce better MCMC mixing properties.

We conducted two analyses.  In the first, we inferred both the species
delimitation and species phylogeny (A11: {\tt speciesdelimitation=1, speciestree=1}).  The posterior probability
is 98.0\% that all the six subspecies are distinct species, with 2\%
probability that M and B are one species.  The best supported
phylogeny is shown in fig.~6, and this has posterior probability 69.2\%.  The
next two trees have different relationships for the three subspecies
of \emph{S. miliarius} (B, M, and S)
from the MAP tree of fig.~6, with posterior probability 21.8\% for (B, (M, S)),
and posterior probability 6.3\% for (M, (B, S)).  Together the three trees have
a cumulative posterior probability of 97.3\% and constitute the 95\% credibility set of models.

In the second analysis (A01: {\tt speciesdelimitation=0, speciestree=1}), we treated the eight species/subspecies as distinct to infer the
species tree.  As in the first analysis, the top three trees differ
concerning the relationships among B, M, and S, with posterior
probability 69.6\% for ((M, B), S) (the MAP tree of fig.~6), 23.3\%
for (B, (M, S), and 6.1\% for (M, (B, S)).
Because the A01 analysis has a reduced model space than the A11
analysis and the shared models in the two analyses are exactly the
same, the posterior probabilities for those shared models should be
proportional in the two analyses.

As the 18 nuclear loci show considerable rate variation \cite[][table 2]{Kubatko:2011}, we repeated
the analysis using a Gamma-Dirichlet model to account for the mutation
rate variation among loci \cite[]{Burgess:2008}.  The gamma parameter in the model is
fixed at $\alpha = 2$.  This has a small effect on the parameter
estimates in the A00 analysis and on the posterior probabilities on
the A11 and A01 analyses.  The results are summarized in table~2.

\noindent
{\bf Analysis of the mitochondrial locus (ATP, 665bp): }  The parameters on
the species tree are assigned the following priors: $\theta \sim G(2, 1000)$ with the prior mean 0.002 and $\tau_0 \sim G(1.5, 10)$ with the
prior mean 0.15.  All other settings are the same as for the
analysis of the nuclear loci.
The mitochondrial locus favoured 5 species, with M and S grouped into
one species in all analyses, in contrast to the nuclear loci, which supported the
distinct species status of all the 6 subspecies (Table~2).  Similarly,
the A01 analysis groups M and S together with posterior probability~97.7\%.
The A00 estimates of species divergence times ($\tau$) are shown in
figure~6c.  The branches in the mitochondrial species tree are much
longer than for the nuclear loci, indicating that the mitochondrial
locus has a much higher mutation rate.  As a
result, the single mitochondrial locus appears to be at least as
informative as the 18 nuclear loci.

The estimated species trees from the nuclear and mitochondrial data have strikingly different shapes.  Relative to the root of the
tree, the mitochondrial species tree has much older nodes for
separation of \emph{S.~catenatus} and \emph{S.~miliarius}, and for the
separation of the the two outgroup species: \emph{A.~contortrix} and
\emph{A.~piscivorus}.  In the simplistic model of random mating and neutral
evolution of both nuclear and mitochondrial loci, the species divergence time parameters  ($\tau$s) should be
proportional for the nuclear and mitochondrial loci.
The ratio of the posterior means of the species divergence times ($\tau$s) between the mitochondrial locus and the nuclear loci is 12 for the root of the species tree, 25 for the common ancestor of \emph{S. catenatus} and \emph{S. miliarius}, and 24 for the divergence of the two outgroup species: \emph{A. contortrix} and \emph{A. piscivorus}.  If the absolute species divergence times are the same for the nuclear and mitochondrial genomes, those estimates indicate that the mitochondrial mutation rate is 12-25 times as high as the nuclear mutation rate.  With such mutation rate differences and if the mitochondrial population size is $\frac{1}{4}$ that for the nuclear loci, we would expect the population size parameters on the species tree ($\theta$s) for the mitochondrial locus to be 3 to 6 times as large as those for the nuclear loci.  Yet, the average of the posterior means for $\theta$s over the populations on the species tree is 0.0019 for the mitochondrial locus and 0.0037 for the nuclear genes, with a ratio of 0.51, while the average of the ratios is 0.81.  Thus the mitochondrial $\theta$s are far smaller than expected from the simple neutral model.
In summary, both the fact that the $\tau$ estimates are not proportional between the nuclear and mitochondrial loci and the fact that the $\tau$ and $\theta$ estimates are not proportional
suggest  that the differences between the nuclear and mitochondrial loci cannot be entirely explained by differences in mutation rates and population sizes alone, and the idealized
model does not fit the data. We suggest that extending the mitochondrial locus and
sequencing more nuclear loci may be useful for understanding the major
factors causing the conflicting signals.

\cite{Kubatko:2011} conducted a number of phylogenetic and coalescent-based analyses of the same data.  The coalescent-based analyses used the heuristic method STEM \cite[]{Kubatko:2009} and the Bayesian MCMC method StarBEAST \cite[]{Heled:2010}.  The 18 nuclear loci and the mitochondrial locus were analyzed as one single dataset.  The species tree inferred in the StarBEAST analysis is the tree shown in figure 6c, with a posterior probability 0.93 for the M-S grouping.  This may be explained by the fact that the mitochondrial locus has a much higher mutation rate so that the signal from the single mitochondrial locus has dominated the analysis when the nuclear and mitochondrial loci are analyzed together.  Note that in our BPP analysis, the relationships among B, M, and S are uncertain, and the mitochondrial locus favors the M-S grouping (fig.~6c and table 2).  Overall our results are largely consistent with the previous analyses of \cite{Kubatko:2011}.

\section*{Discussion}

Our simulation results suggest that the full likelihood-based species tree inference method under the multi-species coalescent model, implemented in BPP, has both high precision (as indicated by the small credibility set) and high accuracy (as indicated by the high coverage probability of the credibility set).  For the parameter combinations examined in our simulation, the correct species tree is recovered with high posterior probabilities when 10 loci are included in the dataset.  The high power of the method is in contrast to the so-called shortcut coalescent methods which use reconstructed gene tree topologies to infer the species tree, ignoring both random and systematic errors in tree reconstruction and ignoring information in the gene tree branch lengths, leading to considerable loss of power.  For example, the MPEST method \cite[]{Liu:2010} produced many weakly supported nodes in the estimated species tree in an analysis of hundreds of loci from the genomes of mammals \cite[]{Song:2012, Gatesy:2013,Wu:2013}.

We note that our current implementation has several limitations.  The first is the use of the \cite{Jukes:1969} mutation/substitution model.  While this appears to be adequate when closely related species are analyzed so that the sequences are highly similar and multiple hits at the same site are uncommon, the model may not be suitable for analysis of distant species such as different orders of mammals or land plants.  It should be straightforward to implement more sophisticated substitution models.  The second is the assumption of the molecular clock, which is expected to be seriously violated in comparison of distantly related species.  It is well-known that clock rooting of phylogenetic
trees is very unreliable when the molecular clock is violated.  Relaxing the molecular clock assumption in the analysis may be more challenging when the sequences are
from multiple individuals of different species related through the multi-species coalescent model .  Note that the relaxed clock models developed for dating species divergences are designed for species data (in which gene trees and species trees are expected to match), and modifications are needed before they can be applied to inference of species phylogenies under the multi-species coalescent model.

\noindent
\textbf{Software availability}
\newline
The algorithms described in this paper are implemented in the program Bayesian Phylogenetics \& Phylogeography (BPP) Version~3.2
which may be downloaded from http://abacus.gene.ucl.ac.uk/software/.

\noindent
\textbf{Acknowledgments}
\newline
This work was supported by a Biotechnological and
Biological Sciences Research Council (UK) grant (to Z.Y.).  Part of
this work was completed when both B.R.\ and Z.Y.\ were guests of Beijing Institute of
Genomics.
\newpage
\bibliographystyle{natbib}
\bibliography{sptree}

\newpage
\begin{landscape}
    \vspace*{\fill}
\begin{center}
\captionof{table}{Summary of results for simulation analyses. }
   \begin{tabular}{lllllllllllllllll}
    \hline
    \multicolumn{1}{c|}{} & \multicolumn{8}{c|}{Number of loci:~2} &  \multicolumn{8}{c|}{Number of loci:~10} \\
    \hline
     \multicolumn{1}{c|}{} & \multicolumn{4}{c|}{Symmetrical tree} &  \multicolumn{4}{c|}{Asymmetrical tree} & \multicolumn{4}{c|}{Symmetrical tree} &  \multicolumn{4}{c|}{Asymmetrical tree} \\
    \hline
     \multicolumn{1}{c|}{} & \multicolumn{2}{c|}{$\theta=0.01$} &  \multicolumn{2}{c|}{$\theta=0.001$} & \multicolumn{2}{c|}{$\theta=0.01$} &  \multicolumn{2}{c|}{$\theta=0.001$}
     & \multicolumn{2}{c|}{$\theta=0.01$} &  \multicolumn{2}{c|}{$\theta=0.001$} & \multicolumn{2}{c|}{$\theta=0.01$} &  \multicolumn{2}{c|}{$\theta=0.001$} \\
    \hline
    Prior & LH & T & LH & T & LH & T & LH & T & LH & T & LH & T & LH & T & LH & T \\ \hline
     \multicolumn{1}{c}{Node} & \multicolumn{16}{c}{Proportion of datasets with true node present in consensus tree} \\
    1           & .98     & .98    & .80    & .76     & .98     & 1.0     & .88     & .86     & 1.0       &        1.0 & 1.0        & 1.0        & 1.0 & 1.0 & 1.0 & 1.0 \\
    2           & 1.0   & .98     & .90   & .84    & .94     & .96      & .88       & .90       & 1.0       & 1.0       & 1.0   & .98     & 1.0 & 1.0 & 1.0 & 1.0 \\
    3           & 1.0   & 1.0   & .88    & .74    & .98     & 1.0   & .84      & .86     & 1.0      & 1.0       & 1.0        & 1.0       & 1.0 & 1.0 & 1.0 & 1.0 \\
    4           & .98  & .94  & .92   & .92    & .96     & 1.0    & .82     & .84      &1.0      & 1.0       & 1.0       & 1.0       & 1.0 & 1.0 & 1.0 & 1.0 \\
    5           & .96  & .92  & .90   & .88   & .94     & .98        & .76      & .80       & 1.0       & 1.0       & 1.0        & 1.0        & 1.0 & 1.0 & 1.0 & 1.0 \\
    6           & 1.0  & 1.0   & .90    & .80     & .96     & .98      & .78      & .84      &1.0        & 1.0       & 1.0        & 1.0        & 1.0 & 1.0 & 1.0 & 1.0 \\
    7           & .96  & .96  & .76    & .64   & .96     & .98       & .76      & .80       & 1.0        & 1.0       & .98        & .98      & 1.0 & 1.0 & 1.0 & 1.0 \\
    8           & 1.0 & .98 & .88   & .82   & .96    & 1.0      & .84     & .86     & 1.0       & 1.0        & 1.0       & 1.0        & 1.0 & 1.0 & .98 & 1.0 \\
    9           & .98 & .98  & .90    & .90    & .94     & .98      & .86     & .88     & 1.0      & 1.0      & 1.0       & 1.0     & 1.0 & 1.0 & 1.0 & 1.0 \\
    10          & .96  & .96    & .80     & .78    & .98     & 1.0      & .86      & .90      & 1.0       & 1.0        & 1.0        & 1.0        & 1.0 & 1.0 & 1.0 & 1.0 \\
    11          & .98   & .94   & .80   & .76    & .96     & 1.0      & .76      & .80       & 1.0      & 1.0        & 1.0        & 1.0        & 1.0 & 1.0 & 1.0 & 1.0 \\
    12          & .98   & .96  & .92   & .88   & .90     & .96      & .68      & .80       &1.0       & 1.0       & 1.0        & 1.0       & 1.0 & 1.0 & 1.0 & 1.0 \\
    13          & .96  & .96  & .76   & .68   & .86     & .94     & .36       & .64      & 1.0      & 1.0        & 1.0        & 1.0       & 1.0 & 1.0 & .98 & 1.0 \\
    14          & .96  & .94   & .78    & .74    & .66     & .80      & .13      & .46      & 1.0      & 1.0       & 1.0        & 1.0      & .94 & .98 & .58 & .80\\
     \multicolumn{1}{c}{CST} & \multicolumn{16}{c}{Empirical coverage} \\
     95 \%   & 1.0     & 1.0     & 1.0     & .98     & 1.0     & 1.0     & .92      & 1.0     & 1.0      & 1.0       & 1.0        & 1.0      & 1.0 & 1.0 & 1.0 & 1.0 \\
     99 \%   & 1.0    & 1.0    & 1.0    & 1.0    & 1.0     & 1.0      & .92   & 1.0    & 1.0      & 1.0       & 1.0        & 1.0      & 1.0 & 1.0 & 1.0 & 1.0 \\
    \multicolumn{1}{c}{CST} & \multicolumn{16}{c}{Mean number of trees in 99\% CST} \\
 & 243.5   & 372.8    & 4375.7    & 6297.3   & 443.9     & 200.9   & 6205.8   & 2711.7   & 1.1      & 1.1       & 15.6        & 19.1     & 3.7 & 3.0 & 46.9 & 26.3 \\
     \hline
    \end{tabular}
\caption*{
The upper matrix shows the proportion of simulated datasets for which each node of the true species tree is present in consensus tre. The empirical coverage of the 95\% and 99\% credible sets of trees (CSTs) tabulates the proportion of simulated datasets (across 50 simulated datasets for each set of simulation conditions) for which the true tree is contained withinin the credible set.  The mean number of trees in the CST is the average number of trees in the 99\% CST (averaging across 50 simulated datasets for each set of simulation conditions). Note -- each dataset is analyzed using
two species tree priors: the uniform prior for labeled histories (LH) and the uniform prior for rooted trees (T). Node numbers are shown in fig.~4.
}
\end{center}
\vspace*{\fill}
\end{landscape}

\newpage
\begin{table}[]
\centering
\caption{Summary of results obtained from BPP analysis of the rattlesnake datasets.}
\begin{tabular}{llll}
\hline
 & 28 nuc loci, one rate & 28 nuc loci, gamma rate G(2) & ATP (665bp) \\\cline{2-4}
 &   $\tau_0 \sim G(1.2,100)$          &  $\tau_0 \sim G(1.2, 100)$            &  $\tau_0 \sim G(1.5, 10)$            \\
 &   $\theta_0  \sim G(2, 1000)$      &  $\theta_0  \sim G(2, 1000)$        &   $\theta_0  \sim G(2, 1000)$         \\
\hline \\
 \textbf{A00 estimates}   &       &                              &             \\
$\tau_0$ (root) & $0.0125~(0.0097, 0.0154)$  &    $0.0139~(0.0100, 0.0172)$   &   $0.145~(0.127, 0.164)$          \\
$\tau_1$ (CET-SMB) & $0.0042~(0.0033, 0.0052)$ & $0.0044~(0.0034, 0.0054)$ & $0.106~(0.088, 0.124)$  \\
$\tau_2$ outgroup  & $0.0026 (0.0013, 0.0039)$ & $0.0026~(0.0014, 0.0039)$ & $0.062~(0.047, 0.077)$      \\ \\
 \textbf{A11 analysis}   &                       &                              &             \\
 Pr(MS) &    $0.000$                   &     $0.000$                         &     $0.968$       \\
 Pr(MB) & $0.020$ & $0.015$ & $0.000$ \\
Pr(ET) & $0.000$ & $0.000$ & $0.335$ \\
$P_4$ & $0.000$ & $0.000$  & $0.330$ \\
$P_5$ & $0.020$ & $0.015$ & $0.652$ \\
$P_6$ & $0.980$ & $0.985$ & $0.018$ \\ \\
 \textbf{A01 analysis}   &                       &                              &             \\
(MB)-S & $0.696$ & $0.608$ & $0.013$ \\
(MS)-B & $0.232$ & $0.313$ & $0.977$ \\
(BS)-M & $0.061$ & $0.071$ & $0.011$ \\ \\
\hline
\end{tabular}
\end{table}

\newpage
\begin{figure}[h]
\centerline{\includegraphics[scale=0.6]{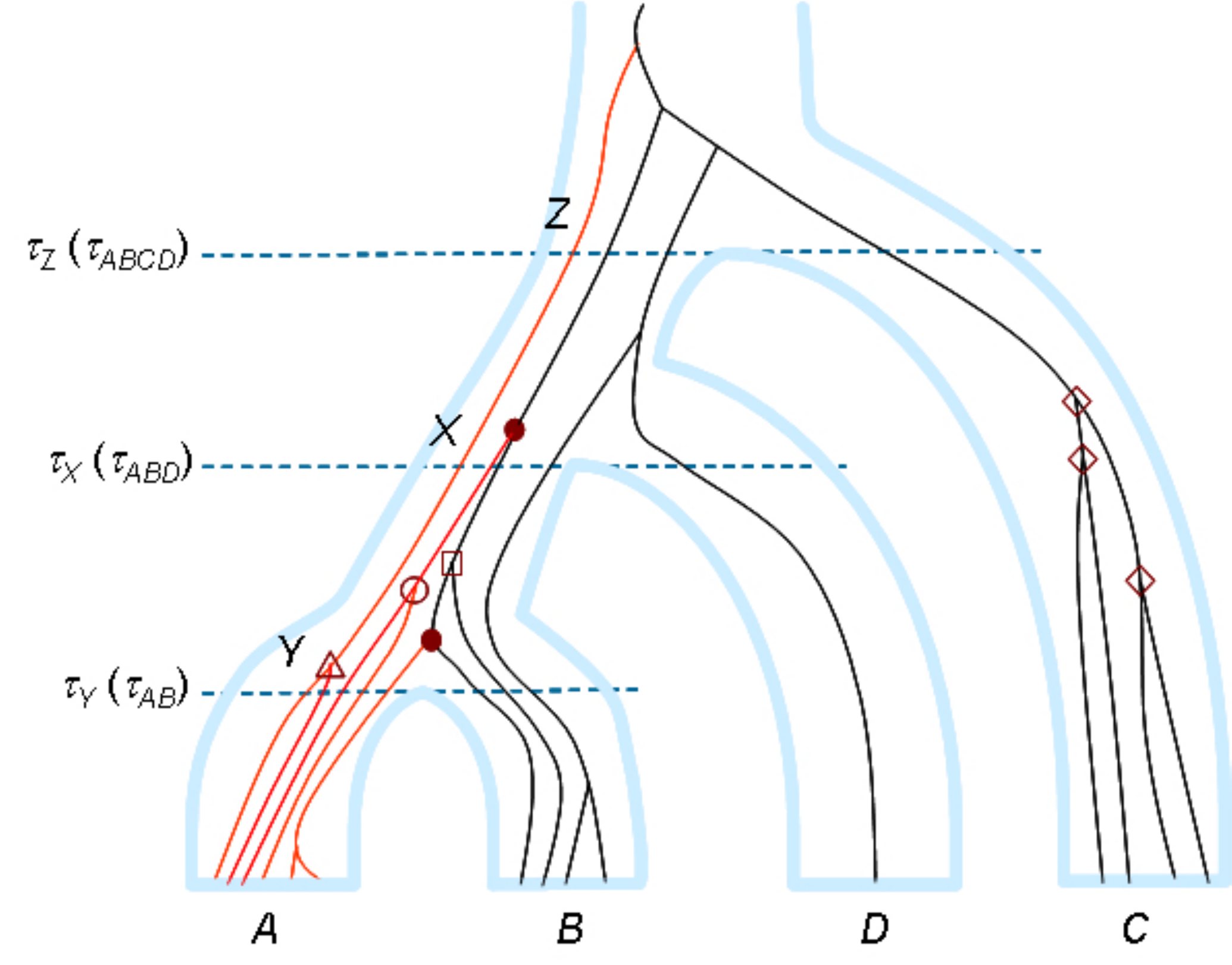}}
\caption{ The SPR move removes branch $Y$-$A$ on the species tree (with the clade $A$)
and reattaches it to a target branch $C$, while changing the gene trees through similar SPR
moves to prevent conflicts between the newly proposed species tree and the gene trees.
Moved nodes on the gene tree, marked as $\bullet$, have exactly one daughter node
with descendents in $A$ only. The species tree is represented by the light blue boundary pipes
while the gene tree is represented by lines contained within the species tree. This is the case with species $AB(Y)$ or the species on the path
from $Y$ to $Z$ (the common ancestor of $A$ and $C$).
Moved nodes are pruned and regrafted to a randomly chosen branch in a species on the path
from $C$ to $Z$. Other affected nodes, marked by $\bigcirc$, $\bigtriangleup$ or $\square$, have their population IDs changed by the move.}\label{SPR}
\end{figure}

\newpage
\begin{figure}[h]
\centerline{\includegraphics[scale=0.6]{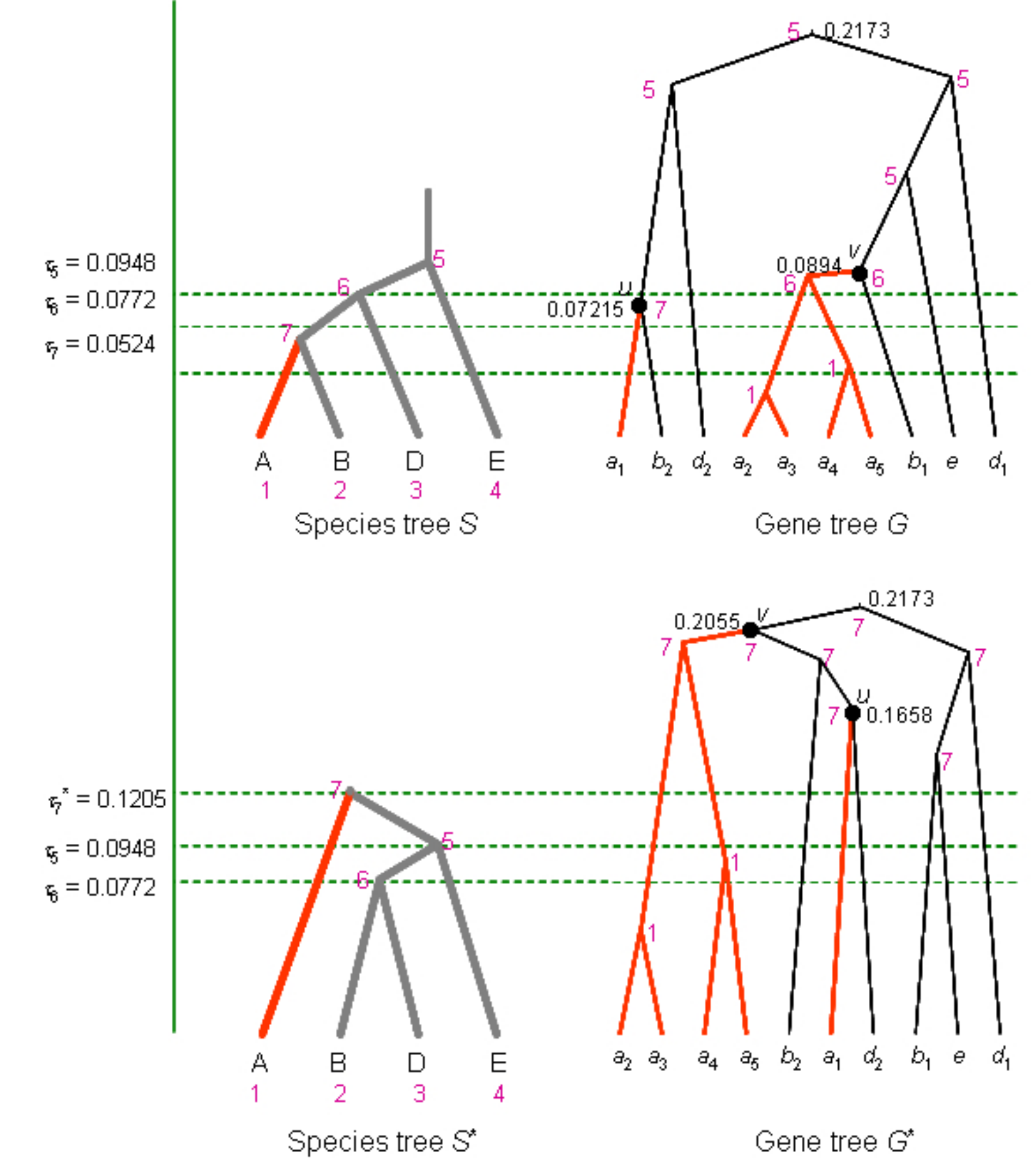}}
\caption{ The nodeslider move prunes branch $Y$-$A$ on the species tree (with clade $A$), changes the age of $Y$ ($\tau_Y$), rescales the ages of
nodes inside $A$ by the factor $\tau^*_Y/\tau_Y$, and reattaches the branch to the species tree either at a branch ancestral to $Y$
(if $\tau^*_Y > \tau_Y$, Expand move) or at a branch that is descendent to $B$  (if $\tau^*_Y < \tau_Y$, Shrink move). Ages of nodes in the affected
clades on the gene trees are scaled by the same factor $\tau^*_Y/\tau_Y$.} \label{NS}
\end{figure}

\newpage
\begin{figure}[h]
\centerline{\includegraphics[scale=0.7]{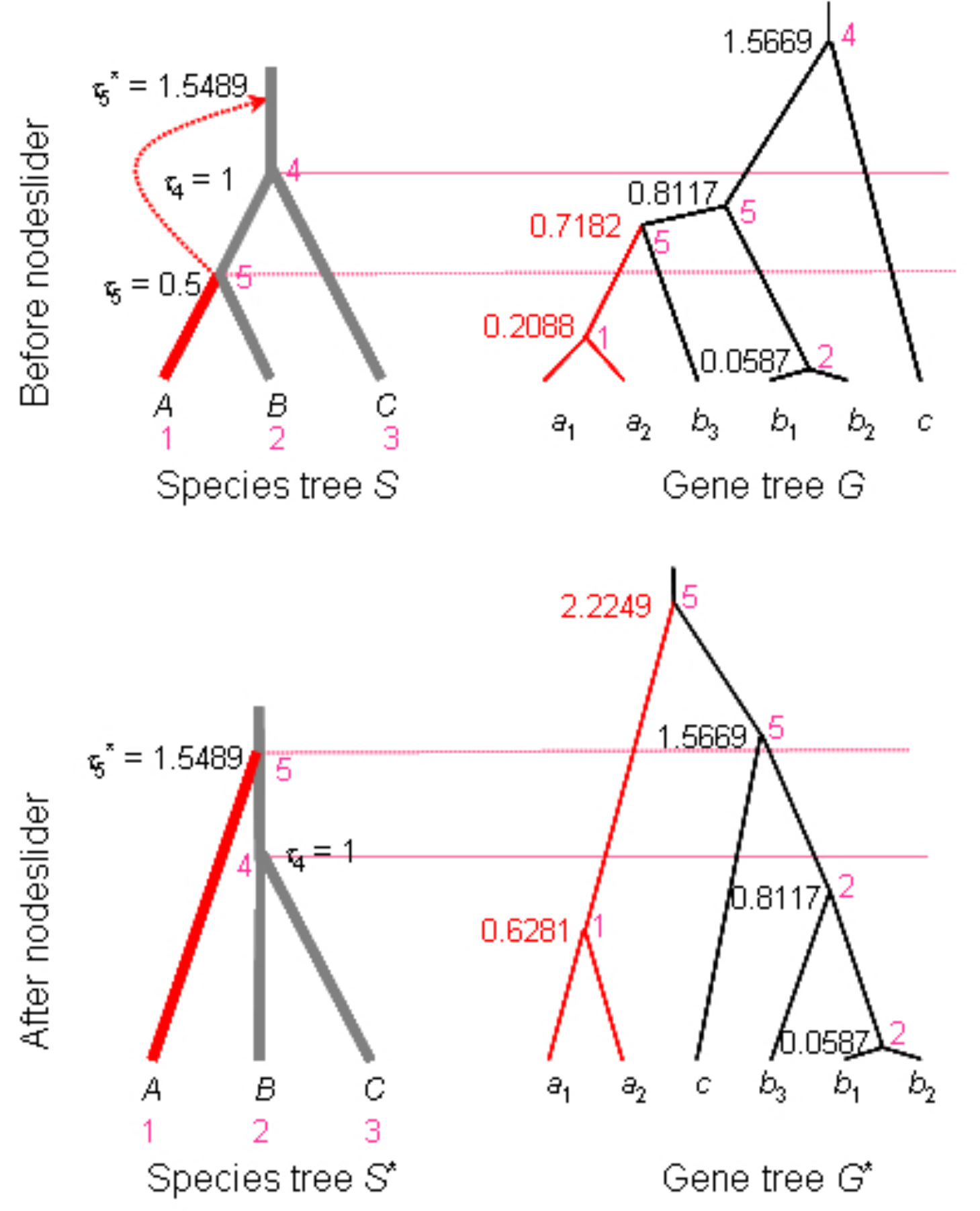}}
\caption{The nodeslider move for 3 species is a variant of the NNI rearrangement for rooted trees, which changes both the species tree topology and a spedcies divergence time ($\tau$),
and always changes the root of the species tree.}\label{3sp}
\end{figure}

\newpage
\begin{figure}[ht]
\centerline{\includegraphics[scale=0.9]{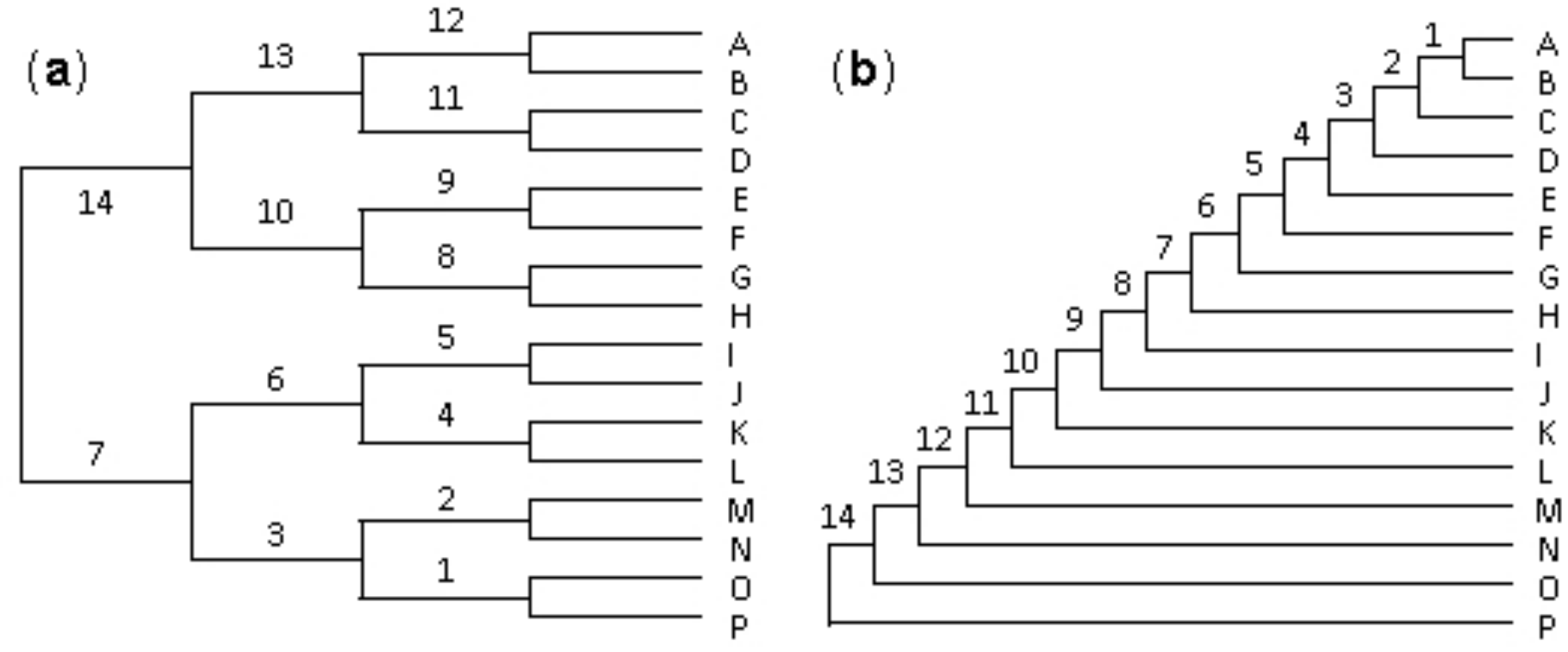}}
\caption{ Symmetrical (a) and asymmetrical (b) species trees used in computer simulation to evaluate the performance of the BPP program.  The branches are
drawn to represent their lengths ($\tau$s) and the 14 nodes are labeled in each tree.}\label{fig4}
\end{figure}

\newpage
\begin{figure}[ht]
\centerline{\includegraphics[scale=0.7]{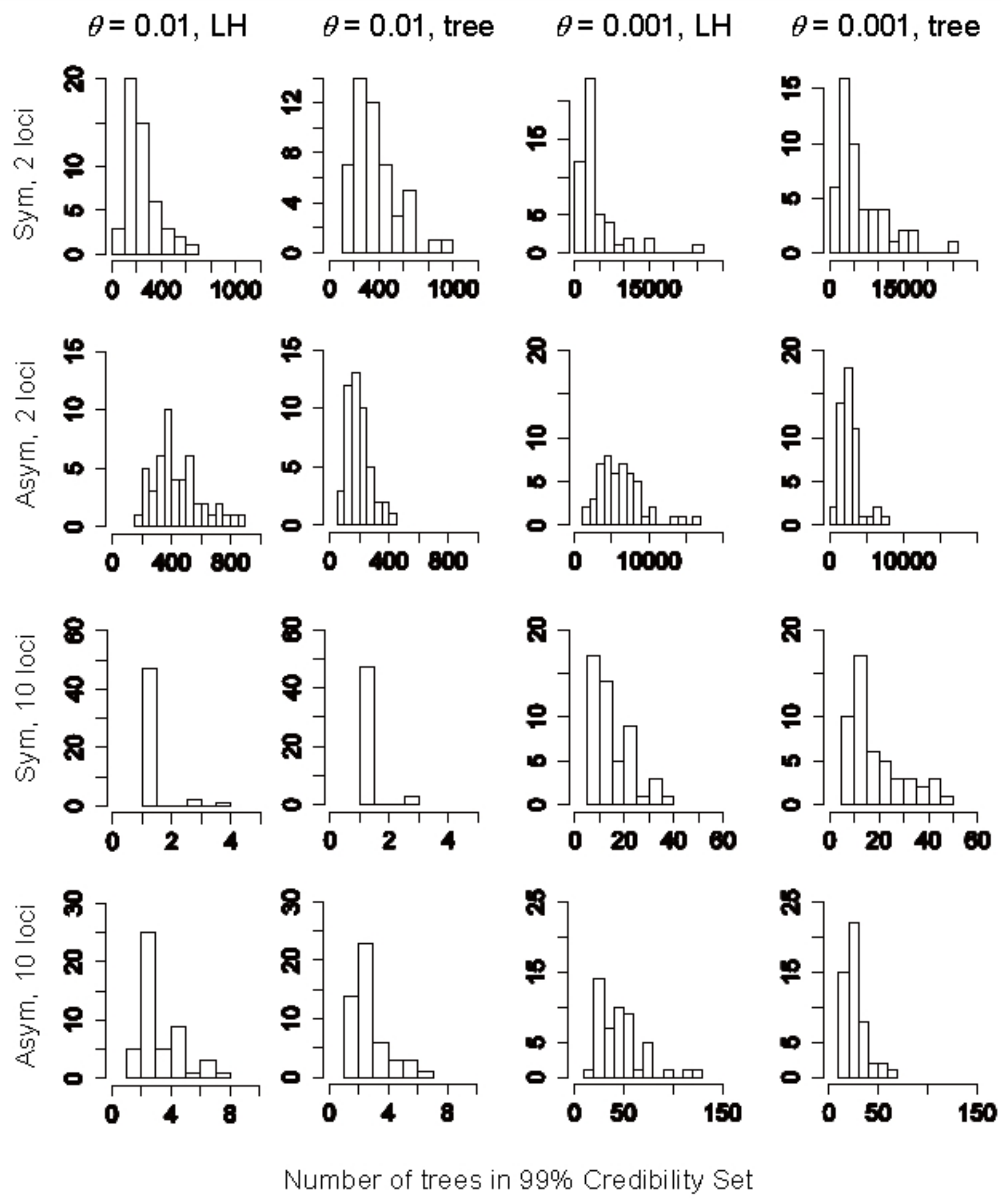}}
\caption{ Frequency distributions of the number of trees contained in the 99\% credible sets from analyses of 50 simulated datasets for each of 8 combinations
of simulation conditions and two different topology priors: LH=labeled history (columns 1 and 3); tree=topology (columns 2 and 4). The upper two rows show results for two loci and and the lower two rows results for 10 loci.
Rows 1 and 3 are results for data simulated on symmetrical (Sym) trees and rows 2 and 4 for asymmetrical (Asym) trees.
The two columns to the left are simulations using $\theta=0.01$ and the two columns on the right are those using $\theta=0.001$} \label{fig5}
\end{figure}

\newpage
\begin{figure}[h]
\centerline{\includegraphics[scale=0.7]{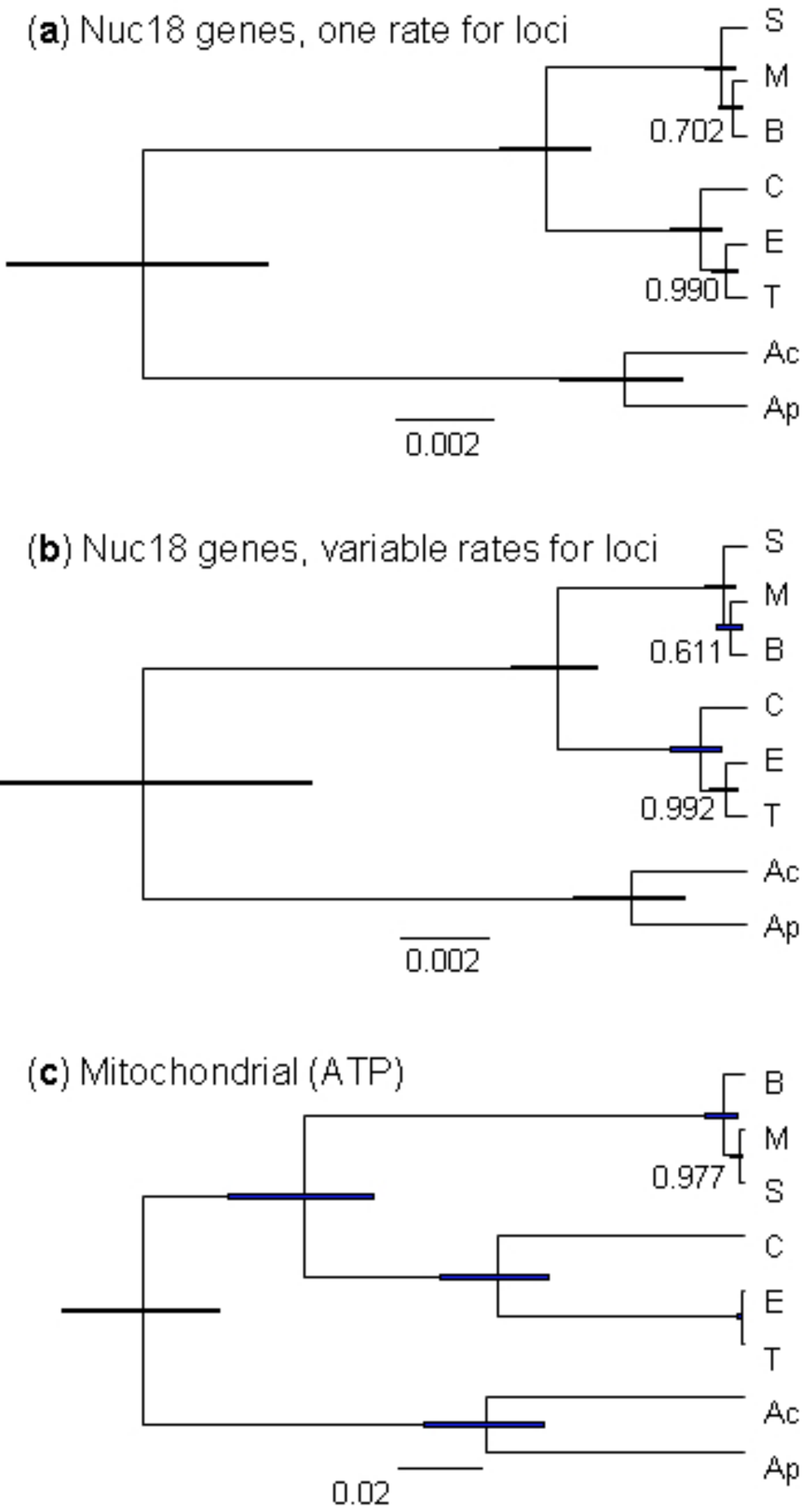}}
\caption{The MAP trees for the six subspecies of \emph{Sistrurus} rattlesnakes and the outgroups in three different analyses of the nuclear (18 loci) and mitochondrial datasets.  The three \emph{S. catenatus} subspecies are \emph{S.~c.~catenatus} (C), \emph{S.~c.~tergeminus} (T), and
\emph{S.~c.~edwardsii} (E), while the three \emph{S.~miliarius}
subspecies are \emph{S.~m.~miliarius} (M), \emph{S.~m.~barbouri} (B),
and \emph{S.~m.~streckeri} (S).   The numbers next to the internal nodes are the posterior probabilities for the clades in the species tree (analysis A01:  {\tt speciesdelimitation = 0, speciestree = 1}).  The branch lengths are drawn to represent the posterior means of the divergence times ($\tau$s) in the A00 analysis ( {\tt speciesdelimitation = 0, speciestree = 0}), with the phylogeny fixed, while the node bars represent the 95\% HPD interval.} \label{F6}
\end{figure}

\end{document}